\documentclass[11pt]{article}

\usepackage{graphicx}
\usepackage{epstopdf}
\usepackage{amsmath}
\usepackage{amssymb}
\usepackage{amsfonts}
\usepackage{amsthm}
\usepackage[usenames]{color}
\usepackage{array}

\usepackage[
colorlinks=true,
linkcolor=blue,
urlcolor=blue,
filecolor=blue,
citecolor=red,
pdfstartview=FitV,
pdftitle={},
pdfauthor={},
pdfsubject={},
pdfkeywords={},
pdfpagemode=None,
bookmarksopen=true
]{hyperref}

\usepackage{epsfig}

\usepackage{hyperref}

\newcommand{\bea}{\begin{eqnarray}}
\newcommand{\eea}{\end{eqnarray}}
\newcommand{\ba}{\begin{array}}
	\newcommand{\ea}{\end{array}}
\newcommand{\ee}{\end{equation}}

\numberwithin{equation}{section}

\begin{document}

\begin{flushright}
	\texttt{\today}
\end{flushright}

\begin{centering}
	
	\vspace{2cm}
	
	\textbf{\Large{
			  Islands in Kerr-de Sitter spacetime and their flat limit  }}
	
	\vspace{0.8cm}
	
	{\large   Sanam Azarnia$^{a}$, Reza Fareghbal$^{a,b}$ }
	
	\vspace{0.5cm}
	
	\begin{minipage}{.9\textwidth}\small
		\begin{center}
			
			{\it  $^{a}$Department of Physics, 
				Shahid Beheshti University, 1983969411,  
				 Tehran , Iran \\
			
			$^{b}$School of Particles and Accelerators, Institute for Research in Fundamental Sciences (IPM)
			P.O. Box 19395-5531, Tehran, Iran}
			
			\vspace{0.5cm}
			{\tt  sanam.azarnia@gmail.com, r$\_$fareghbal@sbu.ac.ir}
			\\ 
			
		\end{center}
	\end{minipage}


	\begin{abstract}
	 We use the quantum extremal island method to study the information paradox in certain cosmological setups known as three dimensional Kerr-de Sitter spacetimes. To do so, we  couple   an auxiliary flat bath system to this spacetime in timelike singularity and measure entropy of Hawking radiation in its asymptotic regions where the gravity is weak. We show that adding the island regions to the entanglement wedge of the radiation causes its entropy to obey  Page curve.  The boundary of the island, i.e., quantum extremal surface is located outside the cosmological horizon in the region connected directly to the bath.  Taking the flat-space limit from the location of the island and its related calculation in the Kerr-de Sitter  sapcetime results in  the  flat-space cosmology (FSC) island, scrambling time and also Page time that were obtained in our previous paper. We repeat the same calculation for the pure de Sitter spacetime and show that our setup which neglects the effect of backreaction, leads also to a quantum extremal surface outside the cosmological horizon. We calculate the scrambling time and confirm the idea that pure de Sitter spacetime is a fast scrambler.

	\end{abstract}

\end{centering}

\newpage

\tableofcontents

\section{INTRODUCTION}
In the past few years there has been a huge renewed effort  to  solve the black hole information paradox by using extremal island method.
According to this  proposal, at late times of radiation some disjoint regions in bulk should be added to the entanglement wedge of the radiation. In other words regions containing the partners of Hawking modes should be considered as a part of the radiation and because of the purification of these pairs, the entanglement entropy starts to decrease and therefore the entropy growth stops.
The boundary of these new regions or islands are  quantum extremal surfaces (QES) and are determined by minimization of generalized entropy for radiation   \cite{Penington:2019npb,Almheiri:2019psf},

\begin{equation}\label{S general}
S_{\text{Rad}}=\text{min}\left(\text{ext}\left[\frac{\text{Area}[\partial I]}{4G_N}+S_{\nu N}(R\cup I)\right]\right).
\end{equation}

In the above relation, the area term refers to the area of the QES and $S_{\nu N}$ is the von Neumann entropy of quantum matter at the union of radiation and island systems in semiclassical theory. The boundary of the island is a codimension-2 object and it is localized both in time and spatial directions. According to the prescription, one has to extremize over the location of island and minimize over all possible islands.
There are multiple examples for the application of the island proposal to different kinds of black hole spacetimes with different asymptotic behavior and dimensions and interestingly in all of them authors have been able to recover the Page curve and namely resolve the information paradox \cite{Almheiri:2019hni,Almheiri:2019yqk,Bousso:2019ykv,Hollowood:2020cou,Gautason:2020tmk,Anegawa:2020ezn,Hartman:2020swn,Almheiri:2019psy,Hashimoto:2020cas,Alishahiha:2020qza,Wang:2021woy,Kim:2021gzd,Ahn:2021chg,Yu:2021cgi,Geng:2021wcq,Saha:2021ohr,Omidi:2021opl}.

  All the questions concerning information paradox in black holes, could be extended to cosmological horizons as they have a similar thermal radiation and one can ask themselves whether or not it is possible to recover the information hidden behind the cosmological horizons. 

As it was shown by Gibbons and Hawking, cosmological event horizon which surrounds the static observer, has thermodynamical properties similar to black hole event horizon  \cite{Gibbons:1977mu}.
If we think of Hawking Radiation as a pair particle-antiparticle creation near the cosmological horizon, the antiparticle crosses the horizon and gets out of reach of the observer, while particles are received by the detector at the center and their flow is defined as radiation coming from cosmological horizon. As it is clear, since the particle-antiparticle pair are highly entangled, the Hawking radiation caries an entanglement entropy that is ever-increasing because of its thermal nature. Obviously, an ever-increasing entropy puts the unitarity of this process in jeopardy, so just like the black hole evaporation case we would like to apply the island conjecture on this problem and investigate the "Page curve" for the radiation initiated from cosmological horizons. 
 
 Recently there have been several investigations regarding the island proposal in cosmological setups (see for example \cite{Sybesma:2020fxg,Balasubramanian:2020xqf,Hartman:2020khs,Aalsma:2021bit,Kames-King:2021etp,Bousso:2022gth,Espindola:2022fqb}). Among them some have focused on the experience of the static observer in pure de Sitter background. In \cite{Aalsma:2021bit} the authors have dimensionally reduced the three dimensional pure de Sitter spacetime to JT gravity on dS without topological term in its action as the so called half reduction and the Nariai black hole to the JT gravity with the topological term as the so called full reduction and have investigated the possibility of information recovery and Page curve for the back-reacted geometry. In the case of half reduction no island was found in the semiclassical regime while for the full reduction island could be found and it was shown that the entropy of radiation follows the Page curve, although it is discussed that due to the impact of the backreaction, a singularity forms and the static observer is doomed to hit it.

In the current letter we would like to investigate the island proposal on the Kerr-de Sitter  (KdS) spacetime \cite{Park:1998qk,Balasubramanian:2001nb} in three dimensions.
This spacetime is particularly interesting for us because not only does it have a cosmological horizon but also at the limit of zero cosmological constant or flat-space limit, one can get to the flat space cosmology (FSC) \cite{Azarnia:2021uch}  in three dimensions which has already been studied from the perspective of island proposal. In \cite{Azarnia:2021uch}, it was shown that in the three dimensional  FSC, information paradox can be solved due to the emergence of an island in late times of radiation. So, it gives us the platform to first study the information recovery for this case and second compare our results with FSC in the flat-space limit of KdS and investigate if there is a well-defined connection or map between quantities that matter in the sense of information recovery. 

In this paper, we use the same method as \cite{Azarnia:2021uch} and couple an auxiliary   flat bath system at timelike singularity of eternal KdS, in order to measure the entropy of radiation at asymptotic regions while keeping this structure eternal. In other words, the energy emitted from KdS is compensated by energy that falls into it from the bath and we can neglect any  gravitational backreaction.

We also use this method for the three dimensional pure de Sitter spacetime which is  different from previous approaches to this problem.

This paper is organized as follows. In sec. II, we introduce the three dimensional Kerr-de Sitter spacetime to some extent. In sec. III, we apply the island conjecture on this spacetime. We calculate  the Page time and scrambling time and also  take the flat-space limit of all results and compare them with FSC. After that, in sec. IV, we turn to the pure de Sitter spacetime.   The last section  is devoted  to the conclusions.
 
\section{THREE DIMENSIONAL KERR-DE SITTER SPACETIME}\label{FSCspacetime}

The Einstein-Hilbert action with positive cosmological constant in three dimensions reads
\begin{equation}
{\it I}=\frac{1}{16 \pi G }\int_{M}d^3x\sqrt{-g}\left(R-\frac{2}{l^2}\right)+\frac{1}{8 \pi G}\int_{\partial M}d^2x\sqrt{\gamma}K\label{action},
\end{equation}
in which $\gamma$ is the induced metric on the boundary and $K$ is the trace of extrinsic curvature.

Using the equations of motion of \eqref{action}, one can find the metric of Kerr-de Sitter (KdS) as,
\begin{equation}
ds^2=-N^2dt^2+N^{-2}dr^2+r^2(d\phi+N^{\phi}dt)^2\label{metric1},
\end{equation}
\begin{equation}
N^2=\mu-\frac{r^2}{l^2}+\frac{16 G^2 J^2}{r^2},\qquad N^{\phi}=\frac{4 G J}{r^2}\label{consvChrg},
\end{equation}
where $\mu$ is a parameter related to mass and $J$ is the angular momentum. Using the  stress tensor  deriving from \eqref{action}, one can find  corresponding conserved charge  related to each Killing vector field in the bulk. The conserved charges associated with time translation and rotational symmetries are \cite{Klemm:2002ir}
\begin{equation}
M:={\it Q}_{\partial_t}=-{{\mu}\over {8 G}}, \qquad {\it Q}_{\partial_\phi}=J.
\end{equation}
 Substituting $\mu$ with $-8GM$ in \eqref{consvChrg} and then solving the equation $N^2=0$, results in a cosmological  horizon  shown by $r_-$, where
\begin{equation}
r_-^2=4 G Ml^2\left(\sqrt{1+\left(\frac{J}{Ml}\right)^2}-1\right).
\end{equation}
It will prove convenient to write   the KdS metric \eqref{metric1} as
\begin{equation}
ds^2=-\frac{(r^2+r_+^2)(r_-^2-r^2)}{l^2 r^2}dt^2+\frac{l^2 r^2}{(r^2+r_+^2)(r_-^2-r^2)}dr^2+r^2\left(d\phi+\frac{r_+r_-}{lr^2}dt\right)^2\label{metric2},
\end{equation} 
where
\begin{equation}
r_+^2= 4G Ml^2\left(\sqrt{1+\left(\frac{J}{Ml}\right)^2}+1\right).
\end{equation}
 The surface gravity of this metric is given by \cite{Park:1998qk}
\begin{equation}\label{kappa for KdS}
\kappa=\frac{r_-^2+r_+^2}{l^2 r_-}.
\end{equation}

Note that the metric \eqref{metric2} and its surface gravity recover the FSC metric and FSC surface gravity \cite{Azarnia:2021uch} in the flat limit, i.e., $r_-\rightarrow r_0$ , $r_+\rightarrow \hat{r}_+l$, and $l\rightarrow \infty$. Moreover,  we should notice that the $\phi$-coordinate is not spacelike everywhere in this geometry. Hence, we define a new periodic coordinate $\psi$ which remains spacelike in the entire spacetime, 
\begin{equation}
\psi=\phi+\frac{r_+}{lr_-}t.
\end{equation}
As a result, our metric at  $\psi=\text{const}$ becomes
\begin{equation}
ds^2=\frac{(r_+^2+r_-^2)(r^2-r_-^2)}{r_-^2 l^2}dt^2-\frac{r^2 l^2}{(r_+^2+r^2)(r^2-r_-^2)}dr^2.\label{null}
\end{equation}

The Penrose diagram of KdS is depicted in Fig. 1.
\begin{figure}[h]
\centering\includegraphics[scale=1,width=70mm]{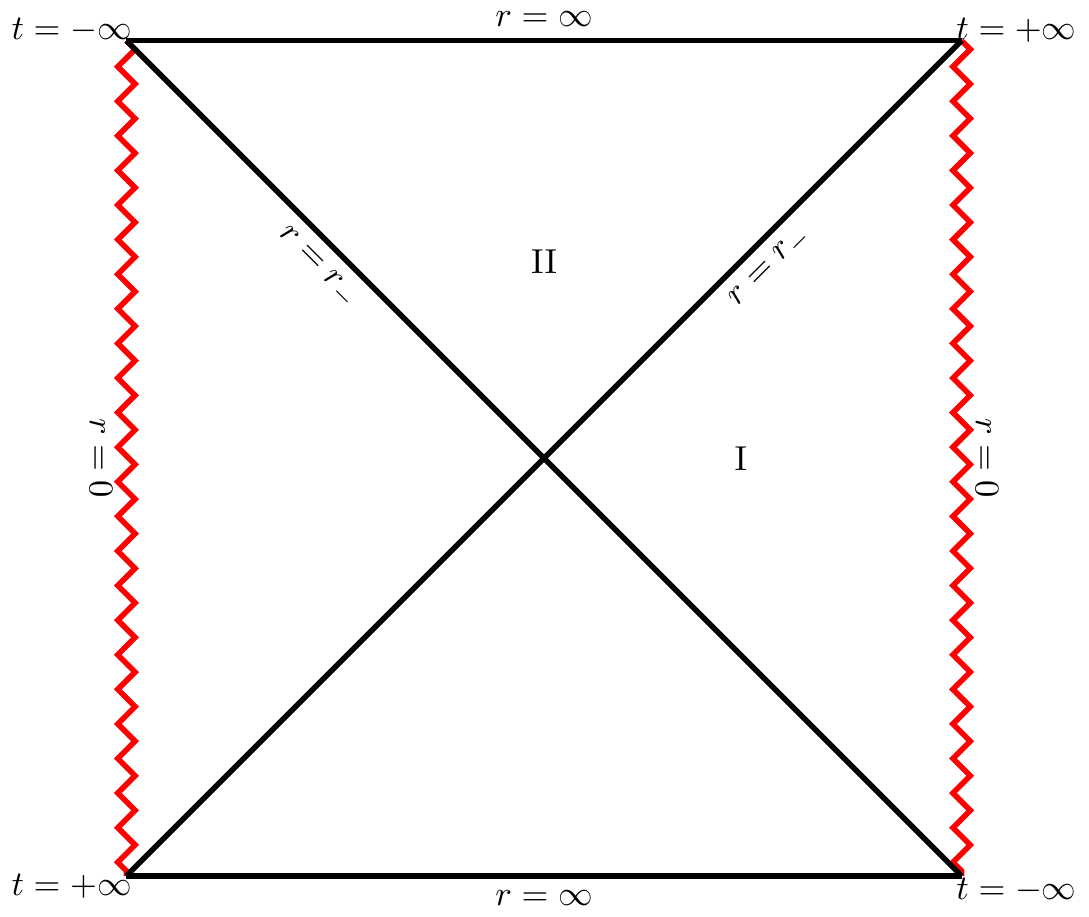}
\caption{The Penrose diagram of KdS. The red wiggly lines represent the timelike singularities while solid lines labeled by $r=r_-$ and $r=\infty$ correspond to the cosmological horizon and future/past infinity respectively}\label{KdS}
\end{figure}

We consider this spacetime to be in thermal equilibrium with a bath at the singularity at the center (Fig. 2) and define the null coordinates as
\begin{equation}\label{null0}
V(t,r)=e^{\kappa\, v(t,r)}=e^{\kappa\, \left(t+r^*(r)\right)},\qquad U(t,r)=-e^{-\kappa\, u(t,r)}=-e^{-\kappa\, \left(t-r^*(r)\right)},
\end{equation}
 where $r^*$ is the tortoise coordinate 
 \begin{equation}
  r^\star_{KdS}(r)=\frac{1}{2\kappa}\log\left[\frac{\sqrt{r_-^2+r_+^2}-\sqrt{r^2+r_+^2}}{\sqrt{r_-^2+r_+^2}+\sqrt{r^2+r_+^2}}4r_+^2\right],\qquad r^*_{Bath}(r)=r.
 \end{equation}
Using the null coordinates \eqref{null0}, the metric will eventually take the following form
\begin{equation}
ds^2=-\Omega^{-2} dUdV \label{dsomega},
\end{equation}
where
\begin{equation}
\Omega_{KdS}(r)=\kappa\frac{lr_-}{\sqrt{(r_-^2-r^2)(r_+^2+r_-^2)}}e^{\kappa r^*_{KdS}}, \qquad \Omega_{Bath}(r)=\kappa e^{\kappa r^*_{\text{Bath}}}.
\end{equation}
and $\kappa$ is given by \eqref{kappa for KdS} for KdS and $\kappa=1$ for bath.
\begin{figure}[h]
\centering\includegraphics[scale=1,width=95mm]{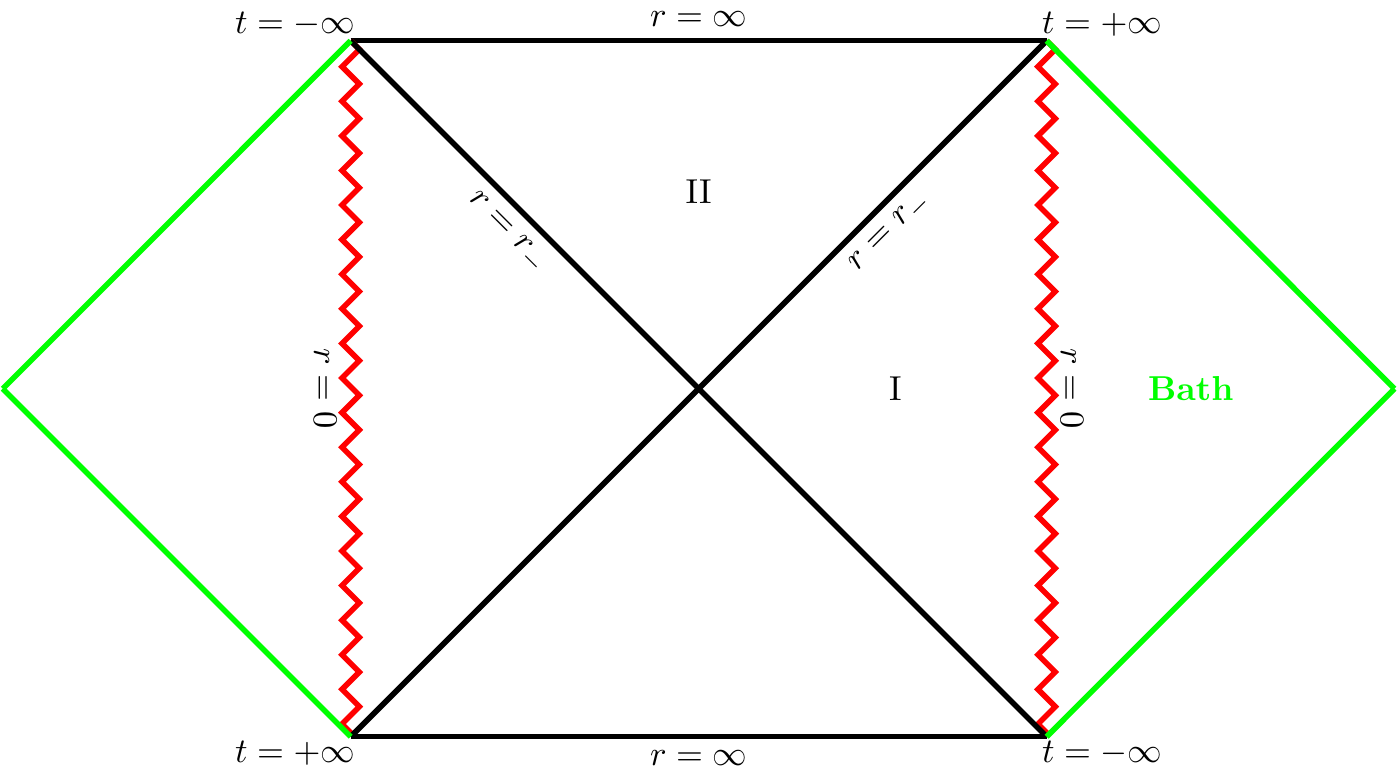}
\caption{The Penrose diagram of KdS+bath.}\label{KdS bath}
\end{figure}

In order to study information paradox in KdS+bath system, we need to consider matter fields which will be exchanged between KdS and bath systems. Thus the full picture requires the matter action $I_{matter}$ to be added to the right hand side of the action\eqref{action}. To simplify the calculations in the next sections, we assume to have conformal matter (to be more precise, free fermion CFT). Having the matter action added to our overall picture can make one worry if the KdS is still a legitimate solution. However, in this paper, similar to \cite{Almheiri:2019yqk}, we study an eternal system and there is no gravitational backreaction because the emitted energy gets completely balanced by the energy that falls in. In other words, the whole system is time-independent and there is no evaporation.

\section{ISLANDS AND THE INFORMATION RECOVERY}\label{islandsection}
Without a detector located at the singularity, every Hawking mode coming from the horizon will be reflected from singularity so the emitted and the absorbed radiation in the horizon are balanced similar to the situation in AdS eternal black holes.

Just as in the FSC setup \cite{Azarnia:2021uch} we connect quantum mechanical systems without gravitational effects as baths to both sides in the Penrose diagram of  KdS. Considering that the system is in thermal equilibrium, this creates a transparent boundary between gravitational and bath systems. 
To get rid of the arealike divergences in the von Newmann entropy of quantum matter, they are absorbed in renormalized Newton constant \cite{Susskind:1994sm} (everywhere throughout this paper $G_N$ represents renormalised Newton constant). In addition we assume that  this cosmological setup is macroscopic and the backreaction is ignored. Hence  the central charge satisfies $1\ll c \ll S_{th}$ and we do not need to consider extra edge modes or extend the Hilbert space. Finally because the entangling regions are far from each other we can use the s-wave approximation at the constant $\psi$ instead of dealing with the more complicated problem of finding entanglement entropy at three dimensional system.

\subsection{Entanglement entropy without islands}
First, we evaluate the entanglement entropy of radiation in the case that islands are absent. In this scenario the gravitational fine grained entropy \eqref{S general} does not get contribution from the area term. We only have regions identified with the Hawking radiation in right and left wedges of the Penrose diagram and the boundaries of these entanglement regions are indicated by $b_+$ and $b_-$ respectively (see Fig. 3).


\begin{figure}[h]
\centering\includegraphics[scale=1,width=90mm]{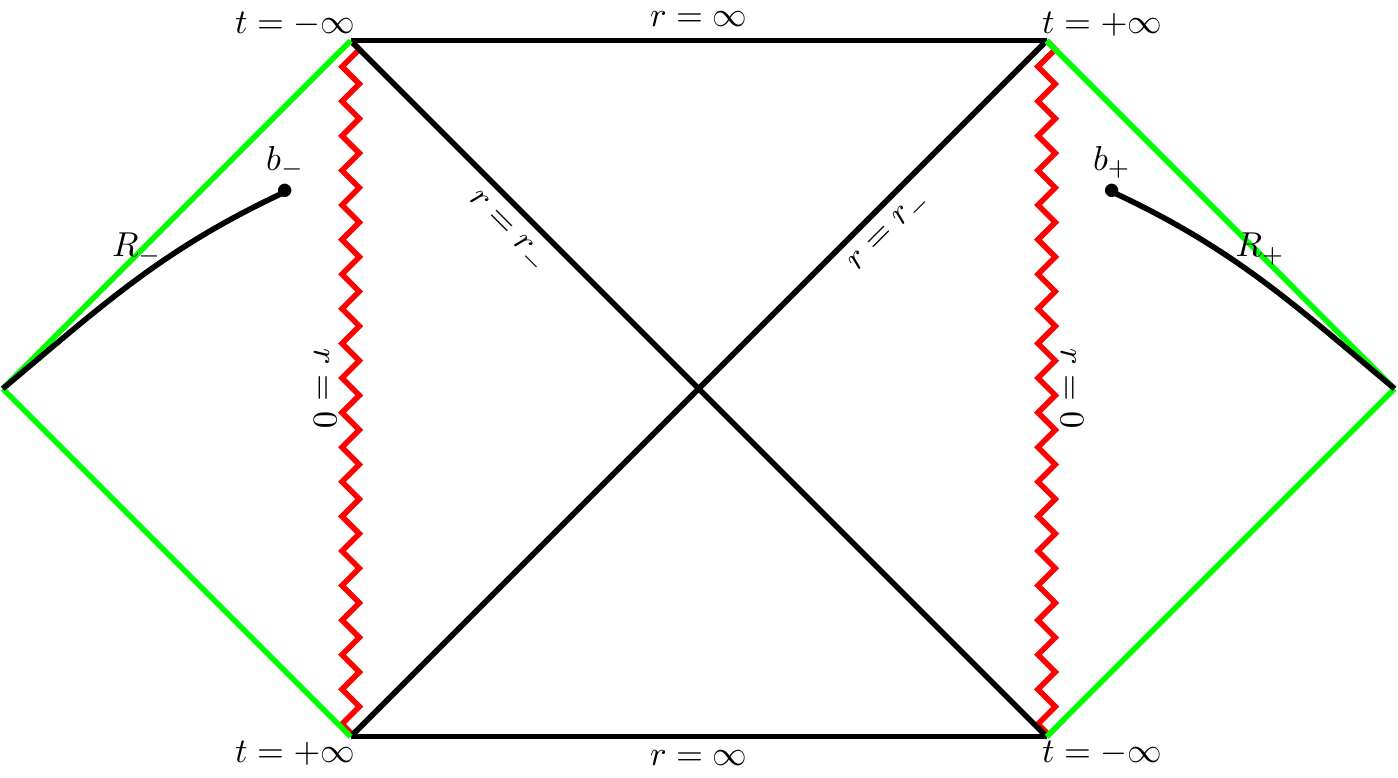}
\caption{$R^+$ and $R^-$ are the regions where the entropy of radiation is measured. In  this case the islands are absent.}\label{KdS}
\end{figure}

 Therefore the generalized entanglement entropy \eqref{S general}  consists only of the  finite quantum matter von Neumann entropy given by
\begin{equation}
S_{gen}=S_{vN,fin}=\frac{c}{6}\log[L(b_+,b_-)]=\frac{c}{6}\log\left[\frac{(V(b_+)-V(b_-))(U(b_-)-U(b_+)}{\Omega(b_+)\Omega(b_-)}\right], \label{no island entropy}
\end{equation}
where $L(b_+,b_-)$ is the geodesic distance between two boundary points $b_+$ and $b_-$.

For the intervals to lie on the same nonzero time slices,  we use the fact that moving  the right side forward and the left side backward in time is an isometry. By implementing this isometry the left and right times can be set equal. Therefore $(t,r)=(t_b,b)$ is set for $b_+$ in right wedge and for $b_-$, $(t,r)=(-t_b+\frac{i \beta}{2},b)$ where $\beta$ is the inverse of temperature.
Hence the entanglement entropy for this case is evaluated as
\begin{equation}
S_{gen}=\frac{c}{3}\log\left[\frac{2}{\kappa} \cosh(\kappa t_b)\right]. \label{growing Ent}
\end{equation} 
This entropy indicates a linear growth at late times,
\begin{equation}
S_{gen}\sim \frac{c}{3}\kappa t_b+... .\label{growing Ent2}
\end{equation} 

This behavior of entanglement entropy at late times is the famous information paradox first noticed by Hawking.
As mentioned before, the island conjecture gives a possibility to solve this paradox by adding extra regions to the entanglement wedge of radiation. These new regions contain the partners of the Hawking modes, so their addition can stop the linear growth of $S_{gen}$ by purifying the state of the emitted particles and their pairs. 

\subsection{Entropy in the presence of islands}\label{3.2}

In order to investigate the presence of the island in the KdS+bath system, we assume that the distance between the right and left wedges is very large so first the s-wave approximation is valid and secondly the von Neumann entropy of the quantum matter in the overall region for the union of the radiation and island is \footnote{ We note that this expression is proved just for free massless fermions (see for example  \cite{Casini:2005rm} and \cite{Casini:2009vk}). In order to use it, we assumed that in competition with the massive Kaluza-Klein  modes, only the massless modes can reach the radiation region. }, 
\begin{equation}
S_{vN,fin}(R\cup I)=\frac{c}{6}\log\left[\frac{L(a_+,a_-)L(b_+,b_-)L(a_+,b_+)L(a_-,b_-)}{L(a_+,b_-)L(a_-,b_+)}\right].
\end{equation}
the assumption of large distances between two wedges implies that
\begin{equation}
L(a_+,a_-)\simeq L(b_+,b_-)\simeq L(a_{\pm},b_{\mp})\gg L(a_{\pm},b_{\pm}).
\end{equation}

Thus, the entanglement entropy of the entire system can be approximated by the sum of the entanglement entropies between the $R\cup I$ in both left and right wedges.(Fig. 4)

\begin{figure}[h]
\centering\includegraphics[scale=1,width=120mm]{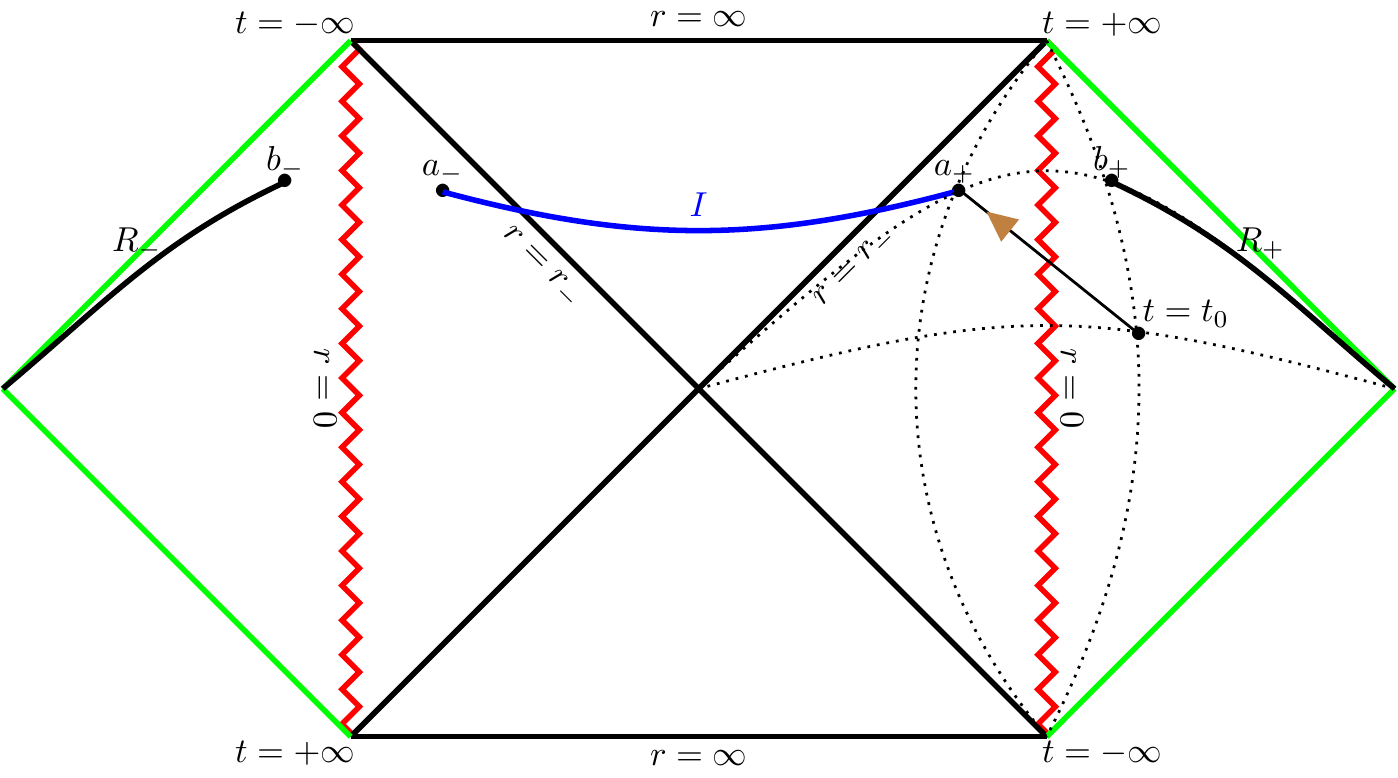}
\caption{KdS+bath system in the presence of island. $a_+$ and $a_-$ are the boundaries of island i.e they are quantum extremal surfaces (QES). $t=t_0$ is the time when a signal is sent toward the island.}\label{KdS1}
\end{figure}


The generalized entropy (for right wedge) will be
\begin{equation}
S_{gen}=\frac{\pi a}{2 G_N}+\frac{c}{6} \log[L(a_+,b_+)]\label{Sgen}
\end{equation}
 Trough calculating the geodesic distance $L(a_+,b_+)$, the generalized entropy is acquired. 
 Extremizing the $S_{gen}$ \eqref{Sgen} with respect to $t_a$ yields 
 \begin{equation}
 {\frac{\partial S_{\text{gen}}}{\partial t_a}=\frac{c \left(e^{2 \kappa t_a}-e^{2 \kappa t_b}\right) \kappa}{6 \left(-e^{\kappa \left(b+t_b\right)}+\frac{2
 e^{\kappa t_a} r_+ \sqrt{-\sqrt{a^2+r_+^2}+\sqrt{r_-^2+r_+^2}}}{\sqrt{\sqrt{a^2+r_+^2}+\sqrt{r_-^2+r_+^2}}}\right) \left(-e^{\kappa \left(-b+t_b\right)}+\frac{e^{\kappa
 t_a} \sqrt{\sqrt{a^2+r_+^2}+\sqrt{r_-^2+r_+^2}}}{2 r_+ \sqrt{-\sqrt{a^2+r_+^2}+\sqrt{r_-^2+r_+^2}}}\right)}}=0.
 \end{equation} 
This equation results in
 \begin{equation}
 t_a=t_b.
 \end{equation}
By implementing this result in $S_{gen}$ we find,
\begin{equation}
S_{\text{gen}}=\frac{\pi  a}{ G_N}+\frac{1}{6} c \log\left[\frac{(r_-^2+r_+^2) (r_-^2-a^2) \left(e^{\kappa b}-2 \sqrt{\frac{r_+^2
\left(\sqrt{r_-^2+r_+^2}-\sqrt{r_+^2+a^2}\right)}{\sqrt{r_-^2+r_+^2}+\sqrt{r_+^2+a^2}}}\right)^4}{4 r_-^2 r_+^2 \kappa^2 l^2 \left(\frac{ \sqrt{r_-^2+r_+^2}-\sqrt{r_+^2+a^2}}{\sqrt{r_-^2+r_+^2}+\sqrt{r_+^2+a^2}}\right)}\right]. \label{gen at island}
\end{equation}
Extremizing this equation with respect to $a$, results in the location of quantum extremal surface to be 
\begin{equation}
a=r_- - \frac{8r_-r_+^2(r_-^2+r_+^2)c^2 e^{2\kappa b} G_N^2}{\left[r_- c(4r_+^2+e^{2\kappa b})G_N+6r_-^2e^{2\kappa b}\pi+6r_+^2e^{2\kappa b}\pi\right]^2}.
\end{equation}
Assuming that the Newton's constant is very small we can estimate the above relation with
\begin{equation}
a_{KdS}=r_- -\frac{2r_- r_+^2 c^2 e^{-2\kappa b }G_N^2}{9(r_-^2+r_+^2)\pi^2}. \label{KdS island}
\end{equation} 
Substituting the location of island \eqref{KdS island} in \eqref{gen at island}, the generalized entropy reads
\begin{equation}\label{Final Sgen}
S_{gen}=2S_{th}+\frac{c}{6}\log\left[\frac{(r_-^2+r_+^2)e^{2\kappa b}}{r_-^2 r_+^2 l^2 \kappa^4}\right]-\frac{2r_- r_+^2 c^2 e^{-2\kappa b }G_N}{3\pi (r_-^2+r_+^2)}.
\end{equation}

Note that  the flat-space  limit of all the above calculations which is implemented by  $r_-\rightarrow r_0$, $r_+\rightarrow l\hat{r}_+$ and $l\rightarrow\infty$ is well-defined and recovers the results of \cite{Azarnia:2021uch}. For example, taking the flat space limit of \eqref{KdS island}  results in
\begin{equation}
a_{FSC}=r_0\left(1-\frac{2c^2 G_N^2 e^{-2\kappa b}}{9\pi^2}\right),
\end{equation}
which is the location of island in FSC.

\subsection{Page time and scrambling time}

Having in mind that the radiation degrees of freedom are encoded in $R\cup I$, the radiation signals that are sent in the KdS will be purified when they reach the island,(see Fig \ref{KdS1}). We are interested in the minimum time interval required for the information to be recovered or the so called ``scrambling time," So considering that the signal thrown into $KdS$ will be added to the radiation degrees of freedom once it reaches the island, we assume that the observer sitting at the radius $b$ sends a signal into $KdS$ at the time $t_0$, (see Fig \ref{KdS1}), this signal will get to the boundary of the island located at the radius $a$ and the time $t_a$. The distance between these two points in the ingoing null direction is the following  

\begin{equation}
v(t_0,b)-v(t_a,a)=(t_0+r^*(b))-(t_a+r^*(a))
\end{equation}

so the time difference between when our signal reaches the boundary of the island and the initial time will be

\begin{equation}
t_a-t_0=r^*(b)-r^*(a)-(v(t_0,b)-v(t_a,a))
\end{equation}

since for the signal to get to the island, $v(t_a,a)$ should be equal or greater than $v(t_0,b)$, the minimum time for the retrieval of information (through the purification process) is given by  

\begin{equation}
t_{scr}\equiv t_a-t_0=r^*(b)-r^*(a).
\end{equation}

 By substituting the location of  island in the above relation we get the following equation for the scrambling time
\begin{equation}
t_{scr(KdS)}=\frac{\beta}{2\pi}\log S_{th}+2b-\frac{\beta}{2\pi}\log \left(\frac{cr_-^2r_+^2}{3(r_-^2+r_+^2)}\right).\label{KdSscr}
\end{equation}
The leading term is universal for fast scramblers and $S_{th}$ is the Bekenstein-Hawking entropy related to the area of KdS horizon at $r=r_-$.  If we take the flat limit again, we will have
\begin{equation}
t_{scr(FSC)}=\frac{\beta}{2\pi}\log S_{th}+2b-\frac{\beta}{2\pi}\log \left(\frac{cr_0^2}{3}\right),
\end{equation}
in which $S_{th}$ is the  entropy coming from the area of  the cosmological horizon $r=r_0$ in FSC. This is exactly the result that we get from direct calculations in FSC \cite{Azarnia:2021uch}.

We can also find the Page time by equating the growing entropy  \eqref{growing Ent} with generalized entropy at the extremal surface which is given by \eqref{Final Sgen}.
So we will have 
\begin{equation}
t_{Page(KdS)}=\frac{3\beta}{\pi c}S_{th}+\frac{\beta}{4\pi}\log\left[\frac{(r_-^2+r_+^2)^2 e^{\frac{4\pi}{\beta} b }\beta^4}{16 \pi^4 r_-^2 r_+^2 l^2}\right]-\frac{\beta r_- r_+^2 c^2 e^{\frac{-4\pi}{\beta} b }G_N}{\pi^2 (r_-^2+r_+^2)}.
\end{equation}
Again, by taking the flat limit we will arrive at  the Page time evaluated for FSC \cite{Azarnia:2021uch}.

\section{COMMENTS ON DE SITTER SPACE}
Now we want to apply the same procedure on the three dimensional pure de Sitter space. Without the islands we will get an ever increasing entropy for the radiation as in \eqref{growing Ent}, \eqref{growing Ent2} \footnote{The surface gravity of de Sitter spacetime is denoted by $\kappa={1\over\ell}$. }. Now we would like to see if we can find an island solution for this spacetime and then look at the problem of information recovery.
\subsection{de Sitter islands, Page time, and scrambling time}
The de Sitter metric in static coordinate and at the constant angular coordinates is
\begin{equation}
ds^2=-\left(1-\frac{r^2}{l^2}\right)dt^2+\left(1-\frac{r^2}{l^2}\right)^{-1} dr^2.
\end{equation}
Again we take this spacetime to be in equilibrium with a thermal bath at the center. (Fig. 5)
\begin{figure}[h]
\centering\includegraphics[scale=1,width=90mm]{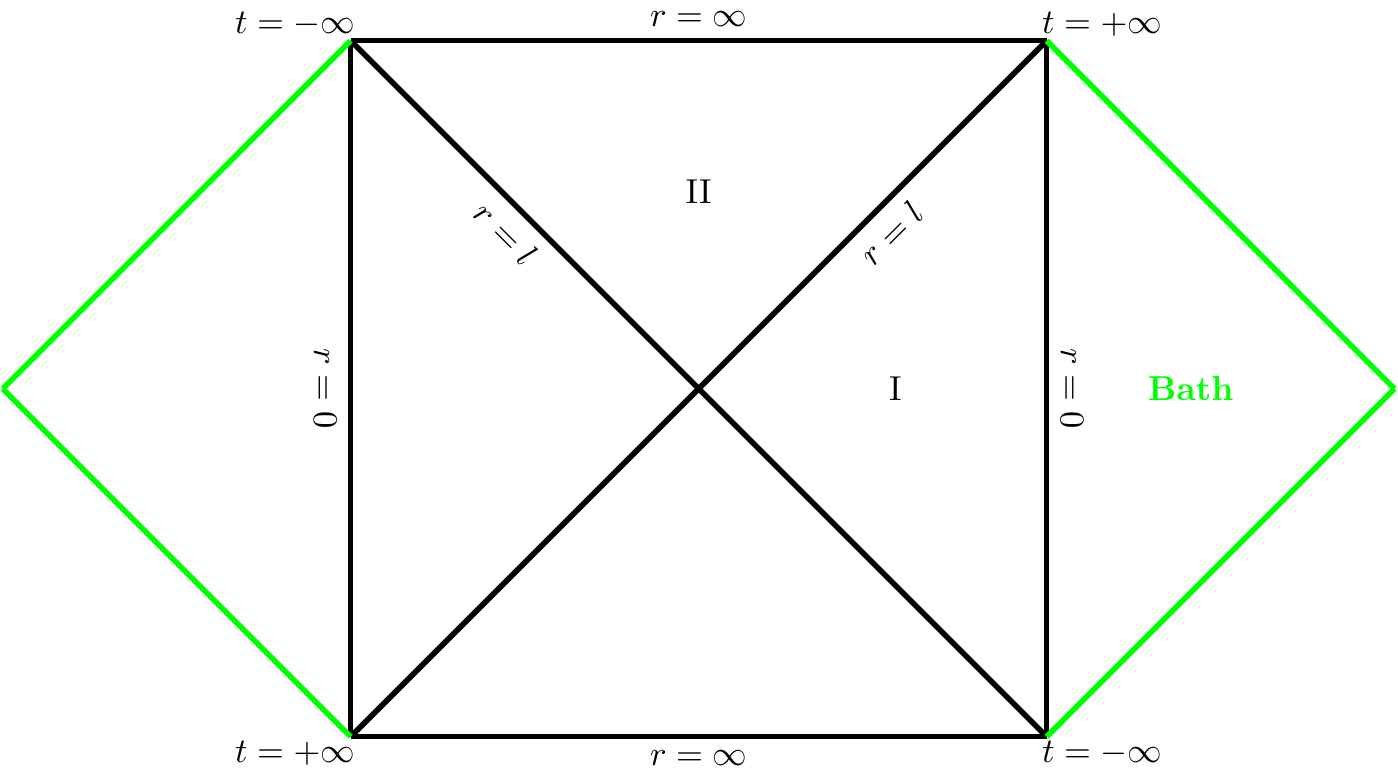}
\caption{The Penrose diagram of de Sitter+bath. }\label{KdS}
\end{figure}
By writing this metric in null coordinates \eqref{null0} and in the form of \eqref{dsomega}, its tortoise coordinate reads
\begin{equation}
r^*_{dS}(r)=\frac{1}{2\kappa}\log\left[\frac{l-r}{l+r}\right], \qquad r^*_{Bath}(r)=r,
\end{equation}
where
\begin{equation}
\Omega_{dS_3}(r)=\frac{\kappa }{\sqrt{1-\frac{r^2}{l^2}}}e^{\kappa\, r^*_{kdS}}, \qquad \Omega_{Bath}(r)=\kappa e^{\kappa\, r^*_{bath}}.
\end{equation}
Following a calculation similar to the calculations of \eqref{3.2} and extremizing \eqref{Sgen} with respect to the location of island, one finds an island solution for this spacetime
\begin{equation}
a_{dS}=l-\frac{c^2 e^{2 \kappa b} G_{N}^2}{18 l \pi^2}. \label{dSisland}
\end{equation}
We should emphasize again that backreaction is not taken into account and it is assumed that the  observer is collecting the radiation in the bath without altering our original spacetime.
Following the same steps as before we can see that the entropy growth stops at the following Page time as the previous case,
\begin{equation}
t_{Page(dS_3)}=\frac{3\beta}{c \pi}S_{th}+\frac{c}{6}\log\left[\frac{e^{\frac{4\pi}{\beta}b}\beta^4}{4\pi^4}\right]-\frac{\beta c e^{-\frac{4\pi}{\beta}b}G_N}{4l\pi^2}.
\end{equation}
Furthermore, the  scrambling time is calculated to be
\begin{equation}
t_{scr(dS_3)}=\frac{\beta}{2\pi}\log{ S_{th}}+2b+\frac{\beta}{2\pi}\log{\frac{12}{c}}.
\end{equation}
  Considering that the leading term is the universal term for the fast scramblers, this result is in agreement with  \cite{Susskind:2011ap} in which it is argued that de Sitter space, like black holes, is a fast scrambler.  It is also in agreement with \cite{Aalsma:2021kle,Geng:2020kxh} where authors have looked at the problem of information exchange in de Sitter spacetime using the shockwave geometry.

 \section{CONCLUSION}
 
In this paper we study the quantum extremal Islands in three dimensional KdS and dS spacetimes. In order to calculate the entropy of radiation, we couple a flat bath system at the timelike singularity of KdS and also at the origin of dS$_3$. In both of these cases we find an island which by using the generalized entropy formula recovers a Page-like curve for the entropy of radiation. 

The boundaries of islands which we find for KdS and dS spacetimes are located outside the cosmological horizon in the region which is connected directly to the bath. This can lead to the causality problem when one decouples the bath from the (K)dS. However, similar to FSC , an unusual formula for the first-law of cosmological horizon as    \cite{Svesko:2022txo}
\begin{equation}\label{FOT}
dE=-TdS,
\end{equation}
can help us to recover causality. In fact, this formula shows that a positive amount of energy which is necessary for decoupling the bath from the (K)dS geometries, reduces the size of horizon and can make the islands to stay completely behind  the horizon.

Another important question in this setup is the way we can extract information from the islands. There is a simple way to do it following \cite{Gao:2016bin}. The large amount of bipartite entanglement between the right and left side of the system ((K)dS+bath) creates a connected geometry, Einstein-Rosen bridge. However, in the presence of this wormhole, a signal from the right-side island cannot reach to the left-side. This means the dual filed theories for the left and right sides do not have any interaction. If we assume the same dictionary  as AdS/CFT, turning on an interaction between left and right dual field theories means that we turn on a nonlocal interaction in the gravity side. Through adjusting the coupling constant of the interaction we can have a shock wave with negative energy in this geometry. Using \eqref{FOT} shows that opposed to the positive energy, propagation of negative energy in the bulk will result in the increase in the size of cosmological horizon and put the island region of right hand side (South pole) in causal connection with the North pole geometry (Fig. 6).
\begin{figure}[h]
\centering\includegraphics[scale=1,width=90mm]{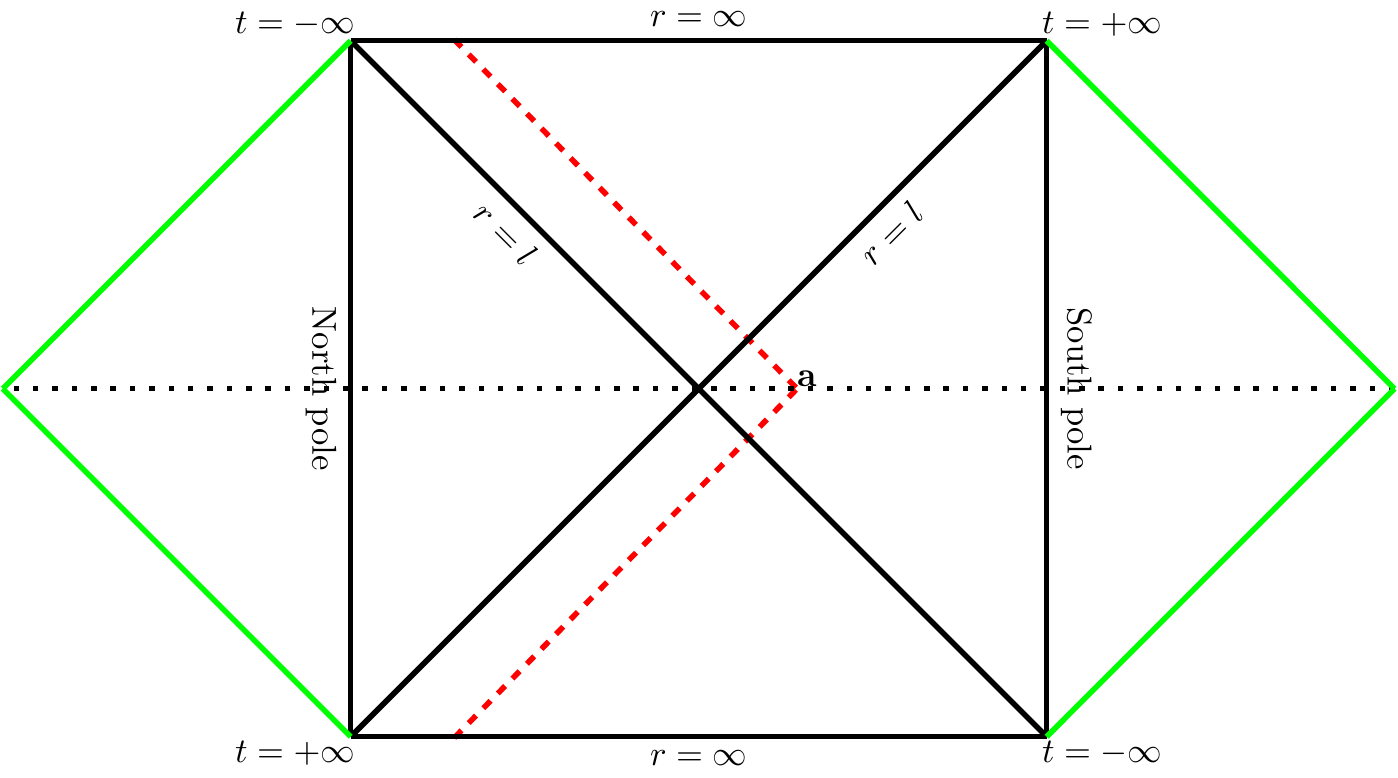}
\caption{A negative energy shock wave can put the left and right side islands in causal connection. }\label{KdS}
\end{figure}
%
  It is important to note that this assessment is only acceptable when backreaction of the negative energy is not taken into account.  
  
  Our main motivation for studying KdS spacetime is its close relation with FSC which is an important geometry in the study of flat-space holography \cite{Bagchi:2012xr}. By taking the flat-space limit from KdS one can find the metric of FSC. In this paper we show that the flat-space limit of the studied quantities in the KdS in the context of quantum extremal islands is well-defined and leads to their related calculations of FSC \cite{Azarnia:2021uch}. This connection between KdS and FSC can be interesting if one uses it to find a connection between flat-space holography and dS holography.

\subsubsection*{Acknowledgments}
The authors would like to thank Ali Naseh for useful comments and fruitful discussions.  This work  is based upon research funded by Iran National Science Foundation (INSF) under Project No. 4003108.

\appendix

\end{document}